\documentclass[aip,jcp,reprint,noshowkeys,superscriptaddress]{revtex4-1}
\usepackage{graphicx,dcolumn,bm,xcolor,microtype,multirow,amscd,amsmath,amssymb,amsfonts,physics,longtable,wrapfig,txfonts,siunitx}
\usepackage[version=4]{mhchem}

\usepackage[utf8]{inputenc}
\usepackage[T1]{fontenc}
\usepackage{txfonts}

\usepackage[
	colorlinks=true,
    citecolor=blue,
    breaklinks=true
	]{hyperref}
\newcommand{\ie}{\textit{i.e.}}

\newcommand{\alert}[1]{\textcolor{black}{#1}}
\usepackage[normalem]{ulem}

\newcommand{\SupMat}{\textcolor{blue}{supplementary material}}

\newcommand{\T}[1]{#1^{\intercal}}

\newcommand{\br}{\boldsymbol{r}}

\newcommand{\dbr}{d\br}

\newcommand{\GW}{\text{$GW$}}

\newcommand{\GOWO}{$G_0W_0$}

\newcommand{\HF}{\text{HF}}
\newcommand{\RPA}{\text{RPA}}

\newcommand{\Ne}{N}
\newcommand{\Norb}{K}
\newcommand{\Nocc}{O}
\newcommand{\Nvir}{V}

\newcommand{\eps}[2]{\epsilon_{#1}^{#2}}
\newcommand{\reps}[2]{\Tilde{\epsilon}_{#1}^{#2}}
\newcommand{\Om}[2]{\Omega_{#1}^{#2}}

\newcommand{\SigC}[1]{\Sigma^\text{c}_{#1}}
\newcommand{\rSigC}[1]{\Tilde{\Sigma}^\text{c}_{#1}}

\newcommand{\MO}[1]{\phi_{#1}}

\newcommand{\ERI}[2]{(#1|#2)}

\newcommand{\bO}{\boldsymbol{0}}
\newcommand{\bI}{\boldsymbol{1}}
\newcommand{\bH}{\boldsymbol{H}}

\newcommand{\bA}[2]{\boldsymbol{A}_{#1}^{#2}}

\newcommand{\bC}[2]{\boldsymbol{C}_{#1}^{#2}}
\newcommand{\bV}[2]{\boldsymbol{V}_{#1}^{#2}}
\newcommand{\bX}[2]{\boldsymbol{X}_{#1}^{#2}}

\newcommand{\bc}[2]{\boldsymbol{c}_{#1}^{#2}}

\newcommand{\RHH}{R_{\ce{H-H}}}
\newcommand{\ii}{\mathrm{i}}

\newcommand{\LCPQ}{Laboratoire de Chimie et Physique Quantiques (UMR 5626), Universit\'e de Toulouse, CNRS, UPS, France}

\begin{document}	

\title{Unphysical Discontinuities, Intruder States and Regularization in $GW$ Methods}

\author{Enzo \surname{Monino}}
	\affiliation{\LCPQ}
\author{Pierre-Fran\c{c}ois \surname{Loos}}
	\email{loos@irsamc.ups-tlse.fr}
	\affiliation{\LCPQ}

\begin{abstract}
By recasting the non-linear frequency-dependent $GW$ quasiparticle equation into a linear eigenvalue problem, we explain the appearance of multiple solutions and unphysical discontinuities in various physical quantities computed within the $GW$ approximation.
Considering the $GW$ self-energy as an effective Hamiltonian, it is shown that these issues are key signatures of strong correlation in the $(N\pm1)$-electron states and can be directly related to the intruder state problem.
A simple and efficient regularization procedure inspired by the similarity renormalization group is proposed to avoid such issues and speed up convergence of partially self-consistent $GW$ calculations.
%\bigskip
%\begin{center}
%	\boxed{\includegraphics[width=0.5\linewidth]{TOC}}
%\end{center}
%\bigskip
\end{abstract}

\maketitle
%%%%%%%%%%%%%%%%%%%%%%%%%%%%%%%%%%%%%%%%%%%%%%%
\section{Introduction}
%%%%%%%%%%%%%%%%%%%%%%%%%%%%%%%%%%%%%%%%%%%%%%%

The $GW$ approximation of many-body perturbation theory \cite{Hedin_1965,Martin_2016} allows to compute accurate charged excitation (\ie, ionization potentials, electron affinities and fundamental gaps) in solids and molecules. \cite{Aryasetiawan_1998,Onida_2002,Reining_2017,Golze_2019}
Its popularity in the molecular electronic structure community is rapidly growing \cite{Ke_2011,Bruneval_2012,Bruneval_2013,Bruneval_2015,Blase_2016,Bruneval_2016, Bruneval_2016a,Koval_2014,Hung_2016,Blase_2018,Boulanger_2014,Li_2017,Hung_2016,Hung_2017,vanSetten_2013,vanSetten_2015,vanSetten_2018, Maggio_2017,vanSetten_2018,Richard_2016,Gallandi_2016,Knight_2016,Dolgounitcheva_2016,Bruneval_2015,Krause_2015,Govoni_2018,Caruso_2016} thanks to its relatively low computational cost \cite{Foerster_2011,Liu_2016,Wilhelm_2018,Forster_2021,Duchemin_2021} and somehow surprising accuracy for weakly-correlated systems. \cite{Korbel_2014,vanSetten_2015,Caruso_2016,Hung_2017,vanSetten_2018,Bruneval_2021}

The idea behind the $GW$ approximation is to recast the many-body problem into a set of non-linear one-body equations. The introduction of the self-energy $\Sigma$ links the non-interacting Green's function $G_0$ to its fully-interacting version $G$ via the following Dyson equation:
\begin{equation}
\label{eq:Dyson}
	G = G_0 + G_0 \Sigma G
\end{equation}
Electron correlation is then explicitly incorporated into one-body quantities via a sequence of self-consistent steps known as Hedin's equations. \cite{Hedin_1965}

In recent studies, \cite{Loos_2018b,Veril_2018,Loos_2020e,Berger_2021,DiSabatino_2021} we discovered that one can observe (unphysical) irregularities and/or discontinuities in the energy surfaces of several key quantities (ionization potential, electron affinity, fundamental and optical gaps, total and correlation energies, as well as excitation energies) even in the weakly-correlated regime. 
These issues were discovered in Ref.~\onlinecite{Loos_2018b} while studying a model two-electron system \cite{Seidl_2007,Loos_2009a,Loos_2009c} and they were further investigated in Ref.~\onlinecite{Veril_2018}, where we provided additional evidences and explanations of these undesirable features in real molecular systems.
In particular, we showed that each branch of the self-energy $\Sigma$ is associated with a distinct quasiparticle solution, and that each switch between solutions implies a significant discontinuity in the quasiparticle energy due to the transfer of weight between two solutions of the quasiparticle equation. \cite{Veril_2018}
Multiple solution issues in $GW$ appears frequently \cite{vanSetten_2015,Maggio_2017,Duchemin_2020} \alert{(even at finite temperature \cite{Pokhilko_2021a,Pokhilko_2021b})}, especially for orbitals that are energetically far from the Fermi level, such as in core ionized states. \cite{Golze_2018,Golze_2020} and finite-temperature scheme. 

In addition to obvious irregularities in potential energy surfaces that hampers the accurate determination of properties such as equilibrium bond lengths and harmonic vibrational frequencies, \cite{Loos_2020e,Berger_2021} one direct consequence of these discontinuities is the difficulty to converge (partially) self-consistent $GW$ calculations as the self-consistent procedure jumps erratically from one solution to the other \alert{even if convergence accelerator techniques such as DIIS \cite{Pulay_1980,Pulay_1982,Veril_2018} or more elaborate schemes  \cite{Pokhilko_2022} are employed.}
Note in passing that the present issues do not only appear in $GW$ as the $T$-matrix \cite{Romaniello_2012,Zhang_2017,Li_2021b,Loos_2022} and second-order Green's function (or second Born) formalisms \cite{SzaboBook,Casida_1989,Casida_1991,Stefanucci_2013,Ortiz_2013,Phillips_2014,Phillips_2015,Rusakov_2014,Rusakov_2016,Hirata_2015,Hirata_2017} exhibit the same drawbacks.

It was shown that these problems can be tamed by using a static Coulomb-hole plus screened-exchange (COHSEX) \cite{Hedin_1965,Hybertsen_1986,Hedin_1999,Bruneval_2006} self-energy \cite{Berger_2021} or by considering a fully self-consistent $GW$ scheme, \cite{Stan_2006,Stan_2009,Rostgaard_2010,Caruso_2012,Caruso_2013,Caruso_2013a,Caruso_2013b,Koval_2014,Wilhelm_2018} where one considers not only the quasiparticle solution but also the satellites at each iteration. \cite{DiSabatino_2021}
However, none of these solutions is completely satisfying as a static approximation of the self-energy can induce significant loss in accuracy and fully self-consistent calculations can be quite challenging in terms of implementation and cost.

In the present article, via an upfolding process of the non-linear $GW$ equation, \cite{Bintrim_2021a} we provide further physical insights into the origin of these discontinuities by highlighting, in particular, the role of intruder states.
Inspired by regularized electronic structure theories, \cite{Lee_2018a,Evangelista_2014b} these new insights allow us to propose a cheap and efficient regularization scheme in order to avoid these issues and speed up convergence of partially self-consistent $GW$ calculations.

Here, for the sake of simplicity, we consider the one-shot {\GOWO} \cite{Strinati_1980,Hybertsen_1985a,Hybertsen_1986,Godby_1988,Linden_1988,Northrup_1991,Blase_1994,Rohlfing_1995,Shishkin_2007} but the same analysis can be performed in the case of (partially) self-consistent schemes such as ev$GW$ \cite{Hybertsen_1986,Shishkin_2007,Blase_2011,Faber_2011,Rangel_2016} (where one updates only the quasiparticle energies) and qs$GW$ \cite{Gui_2018,Faleev_2004,vanSchilfgaarde_2006,Kotani_2007,Ke_2011,Kaplan_2016} (where both quasiparticle energies and orbitals are updated at each iteration). 
Moreover, we consider a Hartree-Fock (HF) starting point but it can be straightforwardly extended to a Kohn-Sham starting point.
Throughout this article, $p$ and $q$ are general (spatial) orbitals, $i$, $j$, $k$, and $l$ denotes occupied orbitals, $a$, $b$, $c$, and $d$ are vacant orbitals, while $m$ labels single excitations $i \to a$.
Atomic units are used throughout.

%%%%%%%%%%%%%%%%%%%%%%%%%%%%%%%%%%%%%%%%%%%%%%%
\section{Downfolding: The non-linear $GW$ problem}
%%%%%%%%%%%%%%%%%%%%%%%%%%%%%%%%%%%%%%%%%%%%%%%

Within the {\GOWO} approximation, in order to obtain the quasiparticle energies and the corresponding satellites, one solve, for each spatial orbital $p$ \alert{and assuming real values of the frequency $\omega$}, the following (non-linear) quasiparticle equation
\begin{equation}
\label{eq:qp_eq}
	\eps{p}{\HF} + \SigC{p}(\omega) - \omega = 0
\end{equation}
where $\eps{p}{\HF}$ is the $p$th HF orbital energy and the correlation part of the {\GOWO} self-energy is constituted by a hole (h) and a particle (p) term as follows
\begin{equation}
\label{eq:SigC}
	\SigC{p}(\omega) 
	= \sum_{im} \frac{2\ERI{pi}{m}^2}{\omega - \eps{i}{\HF} + \Om{m}{\RPA}}
	+ \sum_{am} \frac{2\ERI{pa}{m}^2}{\omega - \eps{a}{\HF} - \Om{m}{\RPA}}
\end{equation}
Within the Tamm-Dancoff approximation (that we enforce here for the sake of simplicity), the screened two-electron integrals are given by
\begin{equation}
	\ERI{pq}{m} = \sum_{ia} \ERI{pq}{ia} X_{ia,m}^\RPA
\end{equation}
where $\Om{m}{\RPA}$ and $\bX{m}{\RPA}$ are respectively the $m$th eigenvalue and eigenvector of the random-phase approximation (RPA) problem, \ie, 
\begin{equation}
	\bA{}{\RPA} \cdot \bX{m}{\RPA} = \Om{m}{\RPA} \bX{m}{\RPA}
\end{equation}
with
\begin{equation}
	A_{ia,jb}^{\RPA} = (\eps{a}{\HF} - \eps{i}{\HF}) \delta_{ij} \delta_{ab} + \ERI{ia}{bj}
\end{equation}
and 
\begin{equation}
	\ERI{pq}{ia} = \iint \MO{p}(\br_1) \MO{q}(\br_1) \frac{1}{\abs{\br_1 - \br_2}} \MO{i}(\br_2) \MO{a}(\br_2) d\br_1 \dbr_2
\end{equation}
are two-electron integrals over the HF (spatial) orbitals $\MO{p}(\br)$. 
Because one must compute all the RPA eigenvalues and eigenvectors to construct the self-energy \eqref{eq:SigC}, the computational cost is $\order*{\Nocc^3 \Nvir^3} = \order*{\Norb^6}$, where $\Nocc$ and $\Nvir$ are the number of occupied and virtual orbitals, respectively, and $\Norb = \Nocc + \Nvir$ is the total number of orbitals.

As a non-linear equation, Eq.~\eqref{eq:qp_eq} has many solutions $\eps{p,s}{\GW}$ \alert{(where the index $s$ is numbering solutions)} and their corresponding weights are given by the value of the following renormalization factor
\begin{equation}
\label{eq:Z}
	0 \le Z_{p,s} = \qty[ 1 - \eval{\pdv{\SigC{p}(\omega)}{\omega}}_{\omega = \eps{p,s}{\GW}} ]^{-1} \le 1
\end{equation} 
In a well-behaved case, one of the solution (the so-called quasiparticle) $\eps{p}{\GW}$ has a large weight $Z_{p}$.
Note that we have the following important conservation rules \cite{Martin_1959,Baym_1961,Baym_1962}
\begin{align}
	\sum_{s} Z_{p,s} & = 1
	&
	\sum_{s} Z_{p,s} \eps{p,s}{\GW} & = \eps{p}{\HF}
\end{align}
which physically shows that the mean-field solution of unit weight is ``scattered'' by the effect of correlation in many solutions of smaller weights.

\alert{In standard $GW$ calculations in solids, \cite{Martin_2016} one assignes a quasparticle peak to the solution of the Dyson equation \eqref{eq:Dyson} that is associated with the largest value of the spectral function
\begin{equation}
	S(\omega) = \frac{1}{\pi} \abs{\Im G(\omega)} 
\end{equation}
}

%%%%%%%%%%%%%%%%%%%%%%%%%%%%%%%%%%%%%%%%%%%%
\section{Upfolding: the linear $GW$ problem}
%%%%%%%%%%%%%%%%%%%%%%%%%%%%%%%%%%%%%%%%%%%%
The non-linear quasiparticle equation \eqref{eq:qp_eq} can be \textit{exactly} transformed into a larger linear problem via an upfolding process where the 2h1p and 2p1h sectors
are upfolded from the 1h and 1p sectors. \cite{Backhouse_2020a,Backhouse_2020b,Bintrim_2021a,Backhouse_2021,Riva_2022}
For each orbital $p$, this yields a linear eigenvalue problem of the form
\begin{equation}
	\bH^{(p)} \cdot \bc{}{(p,s)} = \eps{p,s}{\GW} \bc{}{(p,s)}
\end{equation}
with
\begin{equation}
\label{eq:Hp}
	\bH^{(p)} = 
	\begin{pmatrix}
		\eps{p}{\HF}		&	\bV{p}{\text{2h1p}}	&	\bV{p}{\text{2p1h}}
		\\
		\T{(\bV{p}{\text{2h1p}})}	&	\bC{}{\text{2h1p}}			&	\bO
		\\
		\T{(\bV{p}{\text{2p1h}})}	&	\bO				&	\bC{}{\text{2p1h}}	
	\end{pmatrix}
\end{equation}
where
\begin{align}
	C^\text{2h1p}_{ija,kcl} & = \qty[ \qty( \eps{i}{\HF} + \eps{j}{\HF} - \eps{a}{\HF}) \delta_{jl} \delta_{ac} - 2 \ERI{ja}{cl} ] \delta_{ik} 
	\\
	C^\text{2p1h}_{iab,kcd} & = \qty[ \qty( \eps{a}{\HF} + \eps{b}{\HF} - \eps{i}{\HF}) \delta_{ik} \delta_{ac} + 2 \ERI{ai}{kc} ] \delta_{bd} 
\end{align}
and the corresponding coupling blocks read
\begin{align}
	V^\text{2h1p}_{p,klc} & = \sqrt{2} \ERI{pk}{cl}
	&
	V^\text{2p1h}_{p,kcd} & = \sqrt{2} \ERI{pd}{kc}
\end{align}
The size of this eigenvalue problem is $1 + \Nocc^2 \Nvir + \Nocc \Nvir^2 = \order*{\Norb^3}$, and it has to be solved for each orbital that one wishes to correct.
Thus, this step scales as $\order*{\Norb^9}$ with conventional diagonalization algorithms.
Note, however, that the blocks $\bC{}{\text{2h1p}}$ and $\bC{}{\text{2p1h}}$ do not need to be recomputed for each orbital.
\alert{Of course, this $\order*{\Norb^9}$ scheme is purely illustrative and current state-of-the-art $GW$ implementation scales as $\order*{\Norb^3}$ thanks to efficient contour deformation and density fitting techniques. \cite{Duchemin_2019,Duchemin_2020,Duchemin_2021}}

It is crucial to understand that diagonalizing $\bH^{(p)}$ [see Eq.~\eqref{eq:Hp}] is completely equivalent to solving the quasiparticle equation \eqref{eq:qp_eq}.
This can be further illustrated by expanding the secular equation associated with Eq.~\eqref{eq:Hp}
\begin{equation}
	\det[ \bH^{(p)} - \omega \bI ] = 0 
\end{equation}
and comparing it with Eq.~\eqref{eq:qp_eq} by setting
\begin{multline}
	\SigC{p}(\omega) 
	= \bV{p}{\text{2h1p}} \cdot \qty(\omega \bI - \bC{}{\text{2h1p}} )^{-1} \cdot \T{\qty(\bV{p}{\text{2h1p}})} 
	\\
	+ \bV{p}{\text{2p1h}} \cdot \qty(\omega \bI - \bC{}{\text{2p1h}} )^{-1} \cdot \T{\qty(\bV{p}{\text{2p1h}})} 
\end{multline}
where $\bI$ is the identity matrix.
Because the renormalization factor \eqref{eq:Z} corresponds to the projection of the vector $\bc{}{(p,s)}$ onto the reference (or internal) space, the weight of a solution $(p,s)$ is given by the first coefficient of their corresponding eigenvector $\bc{}{(p,s)}$, \ie, 
\begin{equation}
\label{eq:Z_proj}
	Z_{p,s} = \qty[ c_{1}^{(p,s)} ]^{2}
\end{equation}

One can see this downfolding process as the construction of a frequency-dependent effective Hamiltonian where the internal space is composed by a single Slater determinant of the 1h or 1p sector and the external (or outer) space by all the 2h1p and 2p1h configurations. \cite{Dvorak_2019a,Dvorak_2019b,Bintrim_2021a} 
The main mathematical difference between the two approaches is that, by diagonalizing Eq.~\eqref{eq:Hp}, one has directly access to the internal and external components of the eigenvectors associated with each quasiparticle and satellite, and not only their projection in the reference space as shown by Eq.~\eqref{eq:Z_proj}.

The element $\eps{p}{\HF}$ of $\bH^{(p)}$ [see Eq.~\eqref{eq:Hp}] corresponds to the (approximate) relative energy of the $(\Ne\pm1)$-electron reference determinant (compared to the $\Ne$-electron HF determinant) while the eigenvalues of the blocks $\bC{}{\text{2h1p}}$ and $\bC{}{\text{2p1h}}$, which are $\eps{i}{\HF} - \Om{m}{\RPA}$ and $\eps{a}{\HF} + \Om{m}{\RPA}$ respectively, provide an estimate of the relative energy of the 2h1p and 2p1h determinants.
In some situations, one (or several) of these determinants from the external space may become of similar energy than the reference determinant, resulting in a vanishing denominator in the self-energy \eqref{eq:SigC}.
Hence, these two diabatic electronic configurations may cross and form an avoided crossing, and this outer-space determinant may be labeled as an intruder state.
As we shall see below, discontinuities, which are ubiquitous in molecular systems, arise in such scenarios.

%%%%%%%%%%%%%%%%%%%%%%%%%%%%%%%%%
\section{An illustrative example}
%%%%%%%%%%%%%%%%%%%%%%%%%%%%%%%%%

%%%%%%%%%%%%%%%%%%%%%%%%%%%%%%%%
%	FIGURE 1
%%%%%%%%%%%%%%%%%%%%%%%%%%%%%%%%
\begin{figure}
	\includegraphics[width=\linewidth]{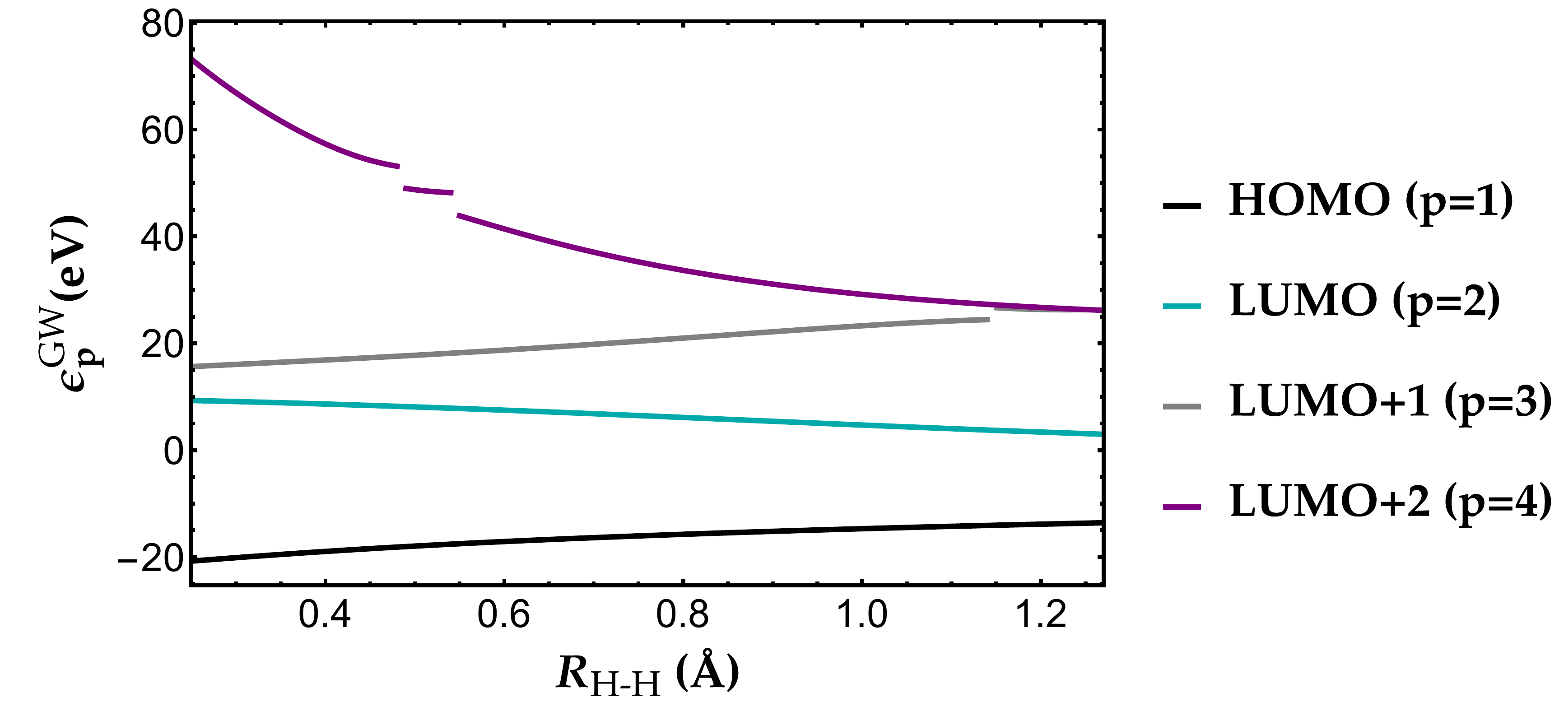}
	\caption{
	\label{fig:H2}
	Quasiparticle energies $\eps{p}{\GW}$ as functions of the internuclear distance $\RHH$ (in \si{\angstrom}) of \ce{H2} at the {\GOWO}@HF/6-31G level.
}
\end{figure}	
%%%%%%%%%%%%%%%%%%%%%%%%%%%%%%%%

%%%%%%%%%%%%%%%%%%%%%%%%%%%%%%%%
%	FIGURE 2
%%%%%%%%%%%%%%%%%%%%%%%%%%%%%%%%
\begin{figure*}
	\includegraphics[width=0.47\linewidth]{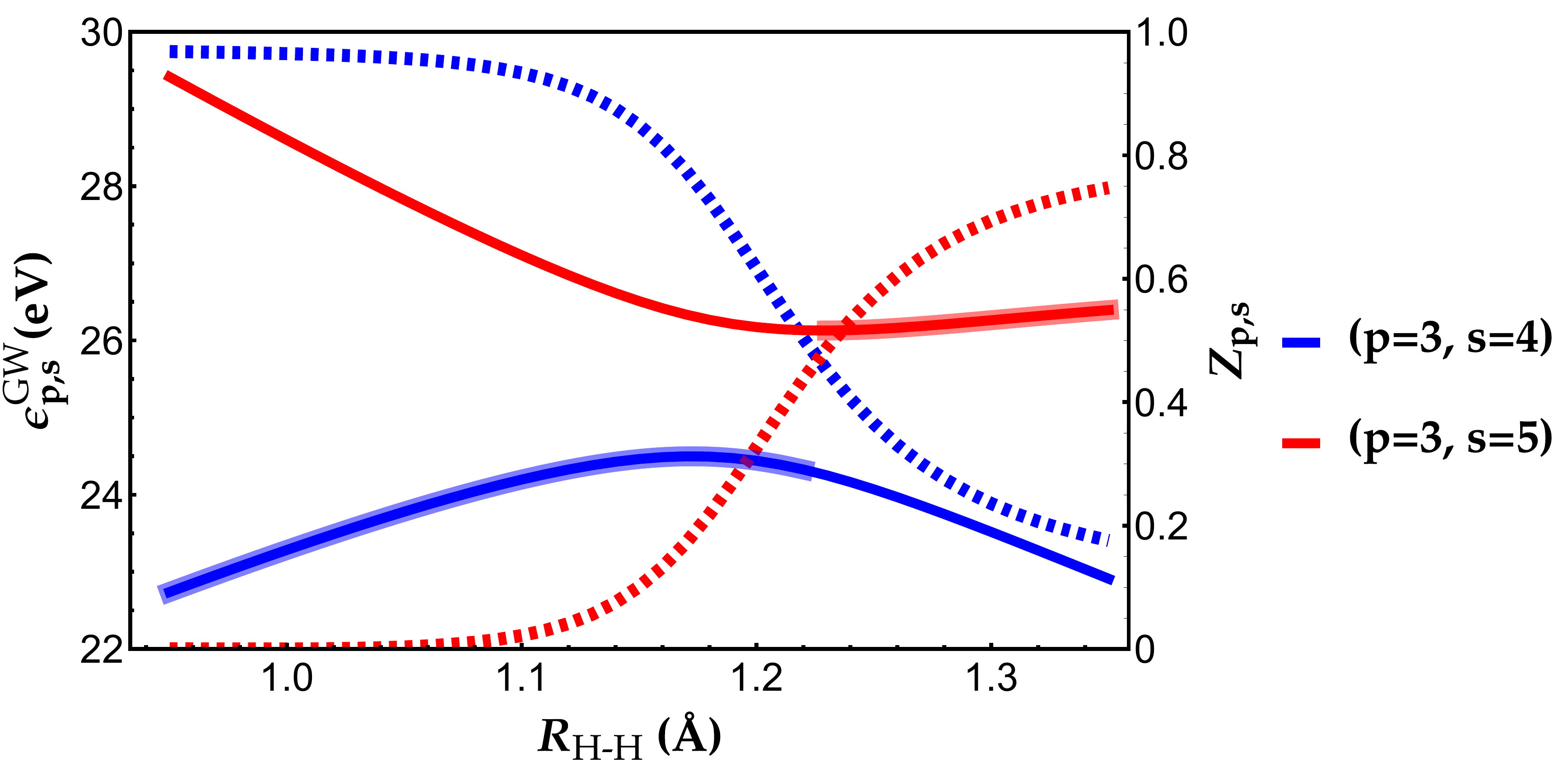}
	\hspace{0.05\linewidth}
	\includegraphics[width=0.47\linewidth]{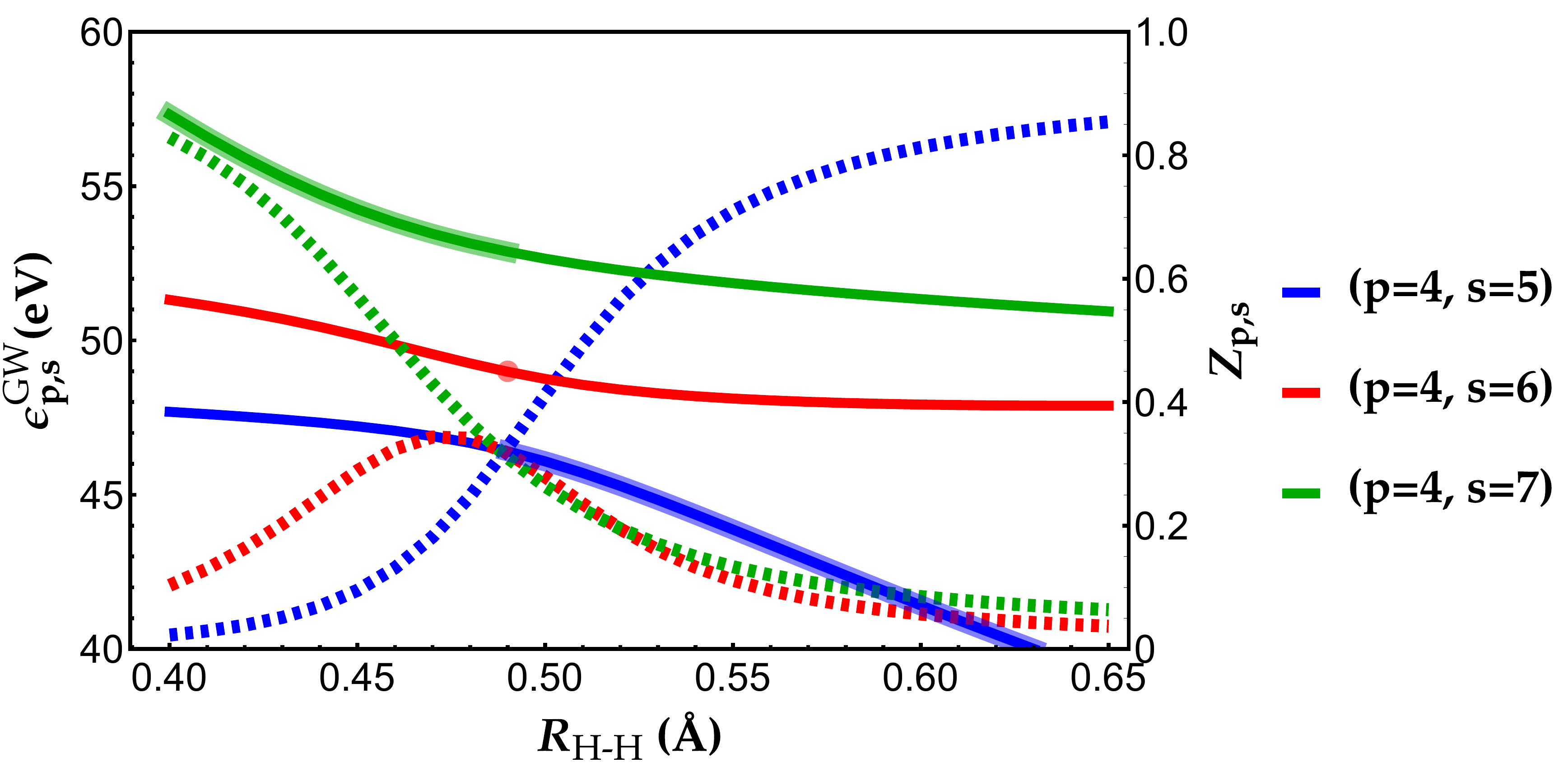}
	\caption{
	\label{fig:H2_zoom}
	Selection of quasiparticle and satellite energies $\eps{p,s}{\GW}$ (solid lines) and their renormalization factor $Z_{p,s}$ (dashed lines) as functions of the internuclear distance $\RHH$ (in \si{\angstrom}) for the LUMO$+1$ ($p=3$) and LUMO$+2$ ($p=4$) orbitals of \ce{H2} at the {\GOWO}@HF/6-31G level.
	The quasiparticle solution (which corresponds to the solution with the largest weight) is represented as a thicker line.
}
\end{figure*}	
%%%%%%%%%%%%%%%%%%%%%%%%%%%%%%%%

In order to illustrate the appearance and the origin of these multiple solutions, we consider the hydrogen molecule in the 6-31G basis set which corresponds to a two-electron system with four spatial orbitals (one occupied and three virtuals).
This example was already considered in our previous work \cite{Veril_2018} but here we provide further insights on the origin of the appearances of these discontinuities.
The downfolded and upfolded {\GOWO} schemes have been implemented in the electronic structure package QuAcK \cite{QuAcK} which is freely available at \url{https://github.com/pfloos/QuAcK}.
These calculations are based on restricted HF eigenvalues and orbitals.
We denote as $\ket*{1\Bar{1}}$ the $\Ne$-electron ground-state Slater determinant where the orbital 1 is occupied by one spin-up and one spin-down electron.
Similar notations will be employed for the $(\Ne\pm1)$-electron configurations.

In Fig.~\ref{fig:H2}, we report the variation of the quasiparticle energies of the four orbitals as functions of the internuclear distance $\RHH$. 
One can easily diagnose two problematic regions showing obvious discontinuities around $\RHH = \SI{1.2}{\angstrom}$ for the LUMO$+1$ ($p = 3$) and $\RHH = \SI{0.5}{\angstrom}$ for the LUMO$+2$ ($p = 4$).
As thoroughly explained in Ref.~\onlinecite{Veril_2018}, if one relies on the linearization of the quasiparticle equation \eqref{eq:qp_eq} to compute the quasiparticle energies, \ie, $\eps{p}{\GW} \approx \eps{p}{\HF} + Z_{p} \SigC{p}(\eps{p}{\HF})$, these discontinuities are transformed into irregularities as the renormalization factor cancels out the singularities of the self-energy. 

Figure \ref{fig:H2_zoom} shows the evolution of the quasiparticle energy, the energetically close-by satellites and their corresponding weights as functions of $\RHH$.
Let us first look more closely at the region around $\RHH = \SI{1.2}{\angstrom}$ involving the LUMO$+1$ (left panel of Fig.~\ref{fig:H2_zoom}).
As one can see, an avoided crossing is formed between two solutions of the quasiparticle equation ($s = 4$ and $s = 5$).
Inspection of their corresponding eigenvectors reveals that the $(\Ne+1)$-electron determinants principally involved are the reference 1p determinant $\ket*{1\Bar{1}3}$ and an excited $(\Ne+1)$-electron determinant of configuration $\ket*{12\Bar{2}}$ that becomes lower in energy than the reference determinant for $\RHH > \SI{1.2}{\angstrom}$.
By construction, the quasiparticle solution diabatically follows the reference determinant $\ket*{1\Bar{1}3}$ through the avoided crossing (thick lines in Fig.~\ref{fig:H2_zoom}) which is precisely the origin of the energetic discontinuity.

A similar scenario is at play in the region around $\RHH = \SI{0.5}{\angstrom}$ for the LUMO$+2$ (right panel of Fig.~\ref{fig:H2_zoom}) but it now involves three solutions ($s = 5$, $s = 6$, and $s = 7$).
The electronic configurations of the Slater determinant involved are the $\ket*{1\Bar{1}4}$ reference determinant as well as two external determinants of configuration $\ket*{1\Bar{2}3}$ and $\ket*{12\Bar{3}}$.
These states form two avoided crossings in rapid successions, which create two discontinuities in the energy surface (see Fig.~\ref{fig:H2}).
In this region, although the ground-state wave function is well described by the $\Ne$-electron HF determinant, a situation that can be safely labeled as single-reference, one can see that the $(\Ne+1)$-electron wave function involves three Slater determinants and can then be labeled as a multi-reference (or strongly-correlated) situation with near-degenerate electronic configurations. 
Therefore, one can conclude that this downfall of $GW$ is a key signature of strong correlation in the $(\Ne\pm1)$-electron states that yields a significant redistribution of weights amongst electronic configurations.

%%%%%%%%%%%%%%%%%%%%%%%%%%%%%%%%%%%%%%%%%%%%%%
\section{Introducing regularized $GW$ methods}
%%%%%%%%%%%%%%%%%%%%%%%%%%%%%%%%%%%%%%%%%%%%%%

%%%%%%%%%%%%%%%%%%%%%%%%%%%%%%%%
%	FIGURE 3
%%%%%%%%%%%%%%%%%%%%%%%%%%%%%%%%
\begin{figure*}
	\includegraphics[width=0.47\linewidth]{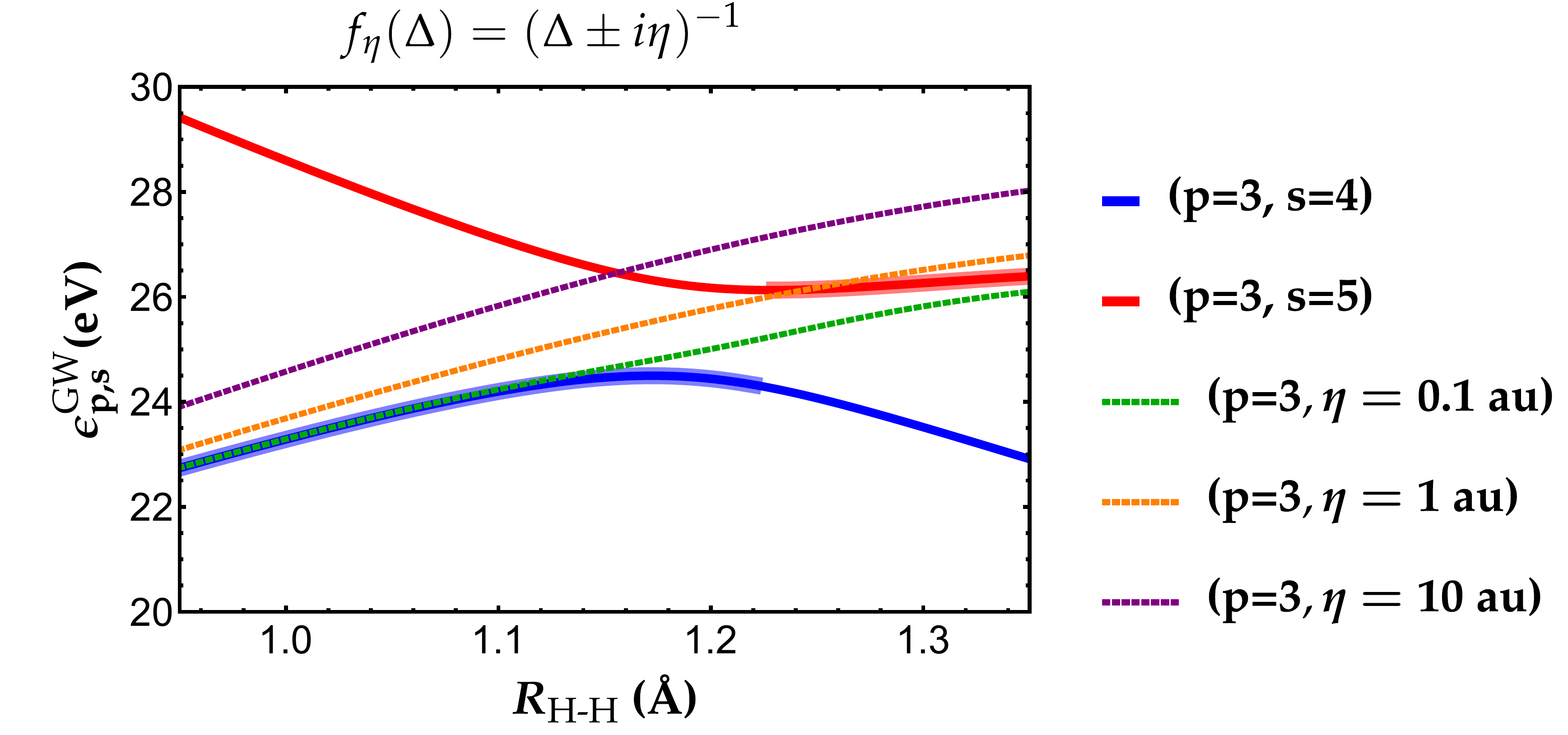}
	\hspace{0.05\linewidth}
	\includegraphics[width=0.47\linewidth]{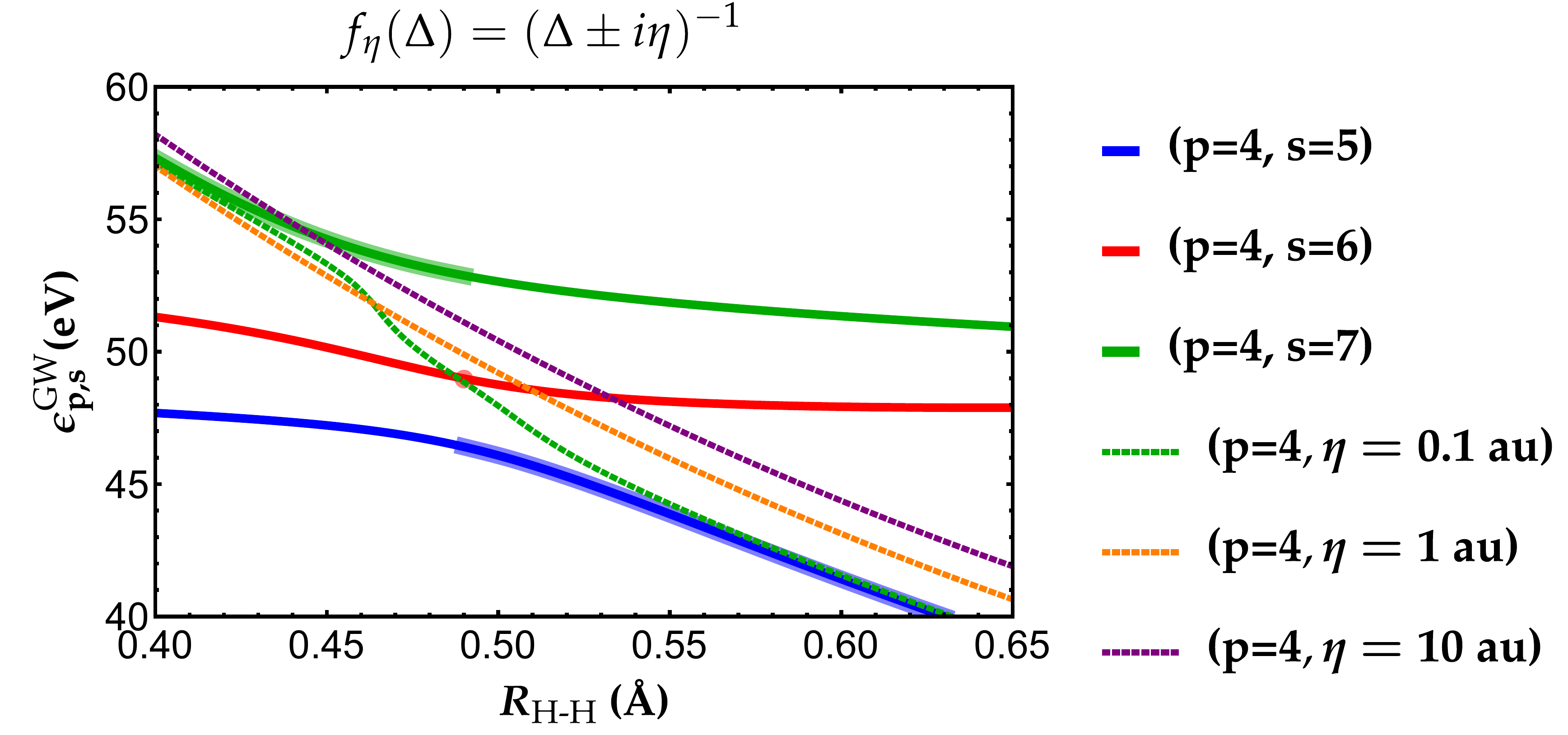}
	\vspace{0.025\linewidth}
	\\
	\includegraphics[width=0.47\linewidth]{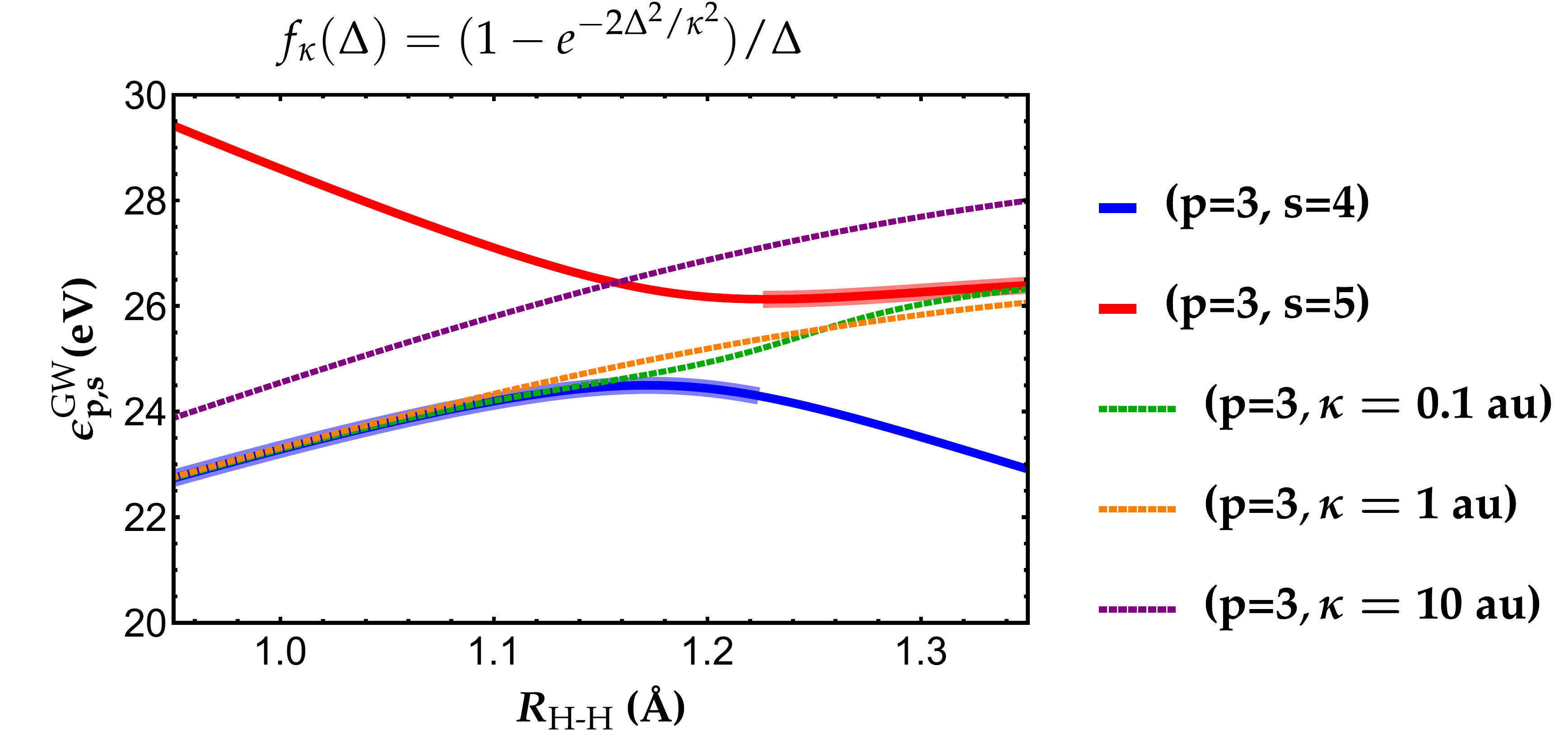}
	\hspace{0.05\linewidth}
	\includegraphics[width=0.47\linewidth]{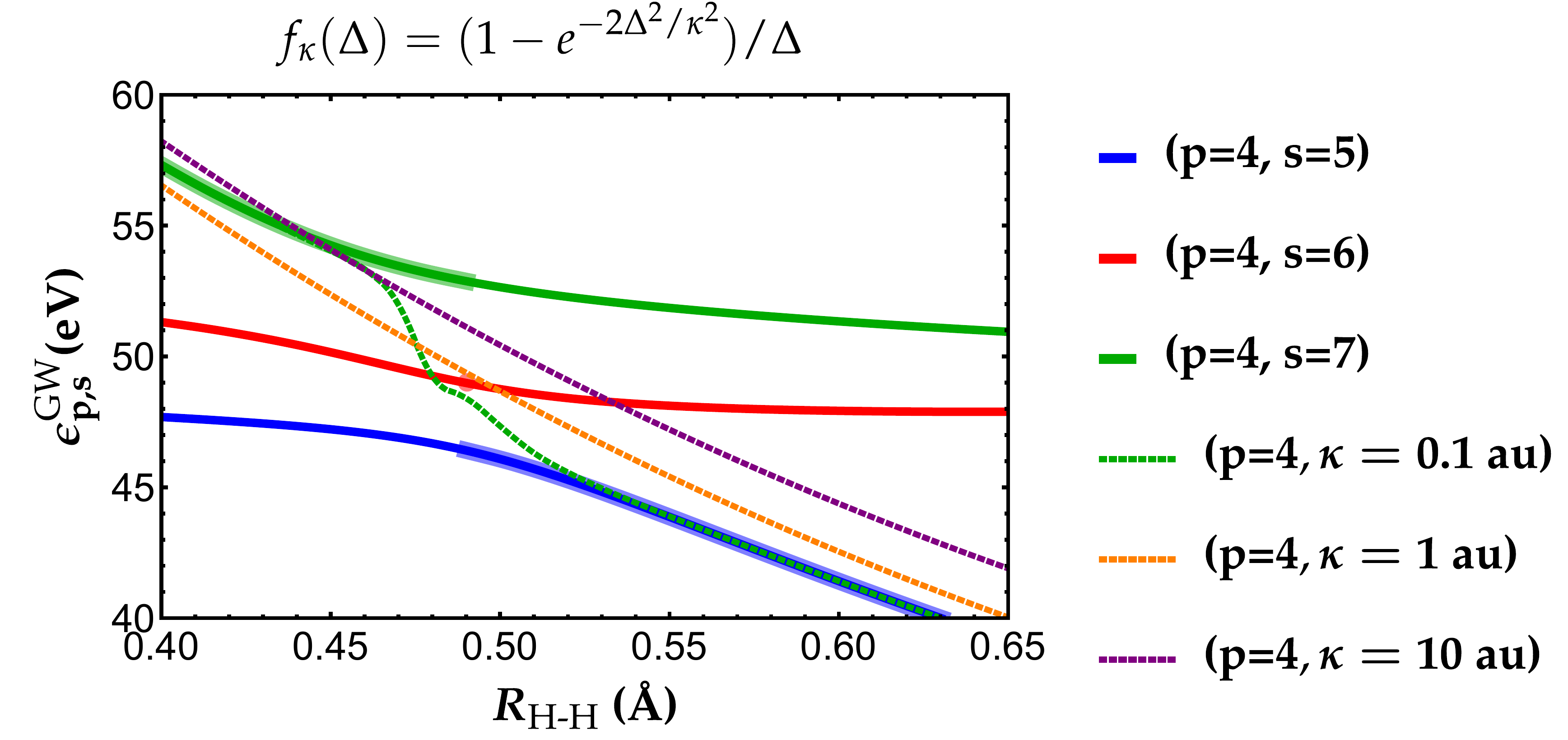}
	\caption{
	\label{fig:H2reg_zoom}
	Comparison between non-regularized (solid lines) and regularized (dashed lines) energies as functions of the internuclear distance $\RHH$ (in \si{\angstrom}) for the LUMO$+1$ ($p=3$) and LUMO$+2$ ($p=4$) orbitals of \ce{H2} at the {\GOWO}@HF/6-31G level.
	The quasiparticle solution is represented as a thicker line.}
\end{figure*}	
%%%%%%%%%%%%%%%%%%%%%%%%%%%%%%%%

%%%%%%%%%%%%%%%%%%%%%%%%%%%%%%%%
%	FIGURE 4
%%%%%%%%%%%%%%%%%%%%%%%%%%%%%%%%
\begin{figure}
	\includegraphics[width=\linewidth]{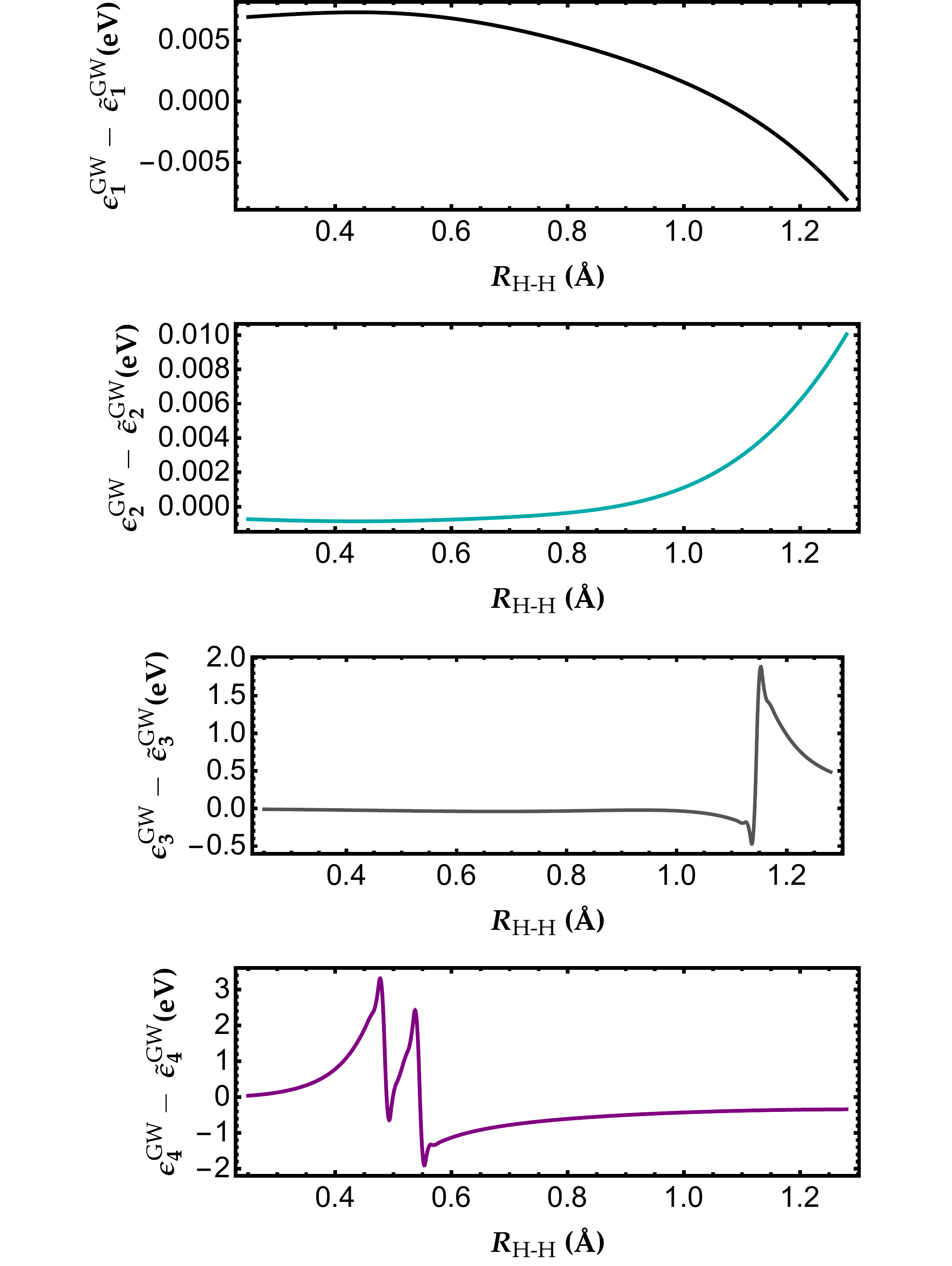}
	\caption{
	\label{fig:H2reg}
	Difference between regularized and non-regularized quasiparticle energies $\reps{p}{\GW} - \eps{p}{\GW}$ computed with $\alert{\kappa} = \SI{1}{\hartree}$ as functions of the internuclear distance $\RHH$ (in \si{\angstrom}) of \ce{H2} at the {\GOWO}@HF/6-31G level.
	\alert{Similar graphs for $\kappa = \SI{0.1}{\hartree}$ and $\kappa = \SI{10}{\hartree}$ are reported as {\SupMat}.}
}
\end{figure}	
%%%%%%%%%%%%%%%%%%%%%%%%%%%%%%%%

%%%%%%%%%%%%%%%%%%%%%%%%%%%%%%%%
%	FIGURE 5
%%%%%%%%%%%%%%%%%%%%%%%%%%%%%%%%
\begin{figure}
	\includegraphics[width=\linewidth]{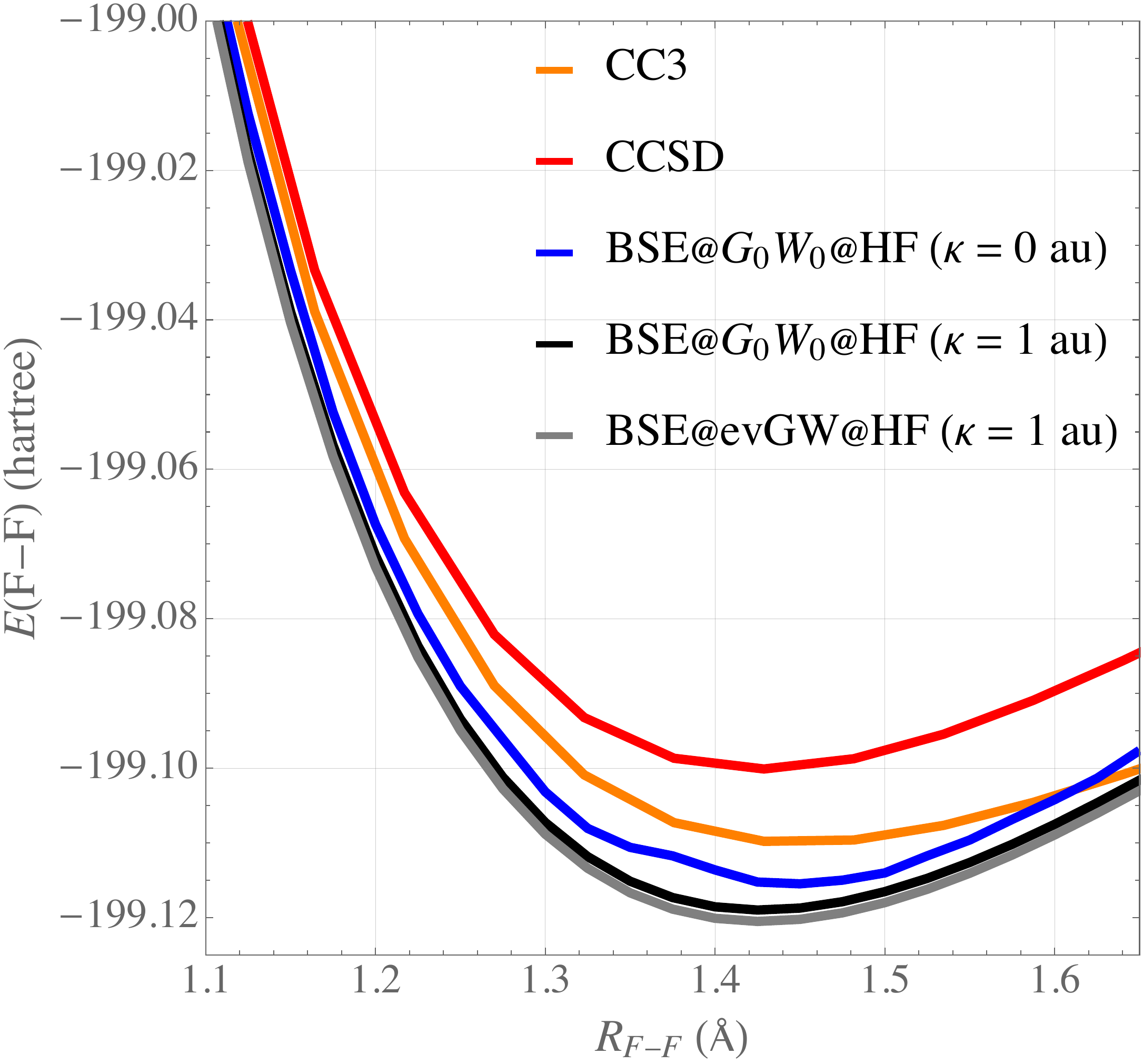}
	\caption{
	\label{fig:F2}
	Ground-state potential energy surface of \ce{F2} around its equilibrium geometry obtained at various levels of theory with the cc-pVDZ basis set.
	\alert{Similar graphs for $\kappa = \SI{0.1}{\hartree}$ and $\kappa = \SI{10}{\hartree}$ are reported as {\SupMat}.}
}
\end{figure}	
%%%%%%%%%%%%%%%%%%%%%%%%%%%%%%%%

One way to alleviate the issues discussed above and to massively improve the convergence properties of self-consistent $GW$ calculations is to resort to a regularization of the self-energy without altering too much the quasiparticle energies.

From a general perspective, a regularized $GW$ self-energy reads 
\begin{equation}
\begin{split}
	\rSigC{p}(\omega;\eta) 
	& = \sum_{im} 2\ERI{pi}{m}^2 f_\eta(\omega - \eps{i}{\HF} + \Om{m}{\RPA})
	\\
	& + \sum_{am} 2\ERI{pa}{m}^2 f_\eta(\omega - \eps{a}{\HF} - \Om{m}{\RPA})
\end{split}
\end{equation}
where various choices for the ``regularizer'' $f_\eta$ are possible.
The main purpose of $f_\eta$ is to ensure that $\rSigC{p}(\omega;\eta)$ remains finite even if one of the denominators goes to zero.
The regularized solutions $\reps{p,s}{\GW}$ are then obtained by solving the following regularized quasiparticle equation:
\begin{equation}
\label{eq:rqp_eq}
	\eps{p}{\HF} + \rSigC{p}(\omega;\eta) - \omega = 0
\end{equation}
Of course, by construction, one must have
\begin{equation}
	\lim_{\eta\to0} \rSigC{p}(\omega;\eta) = \SigC{p}(\omega) 
\end{equation}

The most common and well-established way of regularizing $\Sigma$ is via the simple energy-independent regularizer 
\begin{equation}
\label{eq:simple_reg}
	f_\eta(\Delta) = (\Delta \pm \ii \eta)^{-1}
\end{equation}
(with $\eta > 0$), \cite{vanSetten_2013,Bruneval_2016a,Martin_2016,Duchemin_2020} a strategy somehow related to the imaginary shift used in multiconfigurational perturbation theory. \cite{Forsberg_1997}
\alert{Note that this type of broadening is customary in solid-state calculations, hence such regularization is naturally captured in many codes. \cite{Martin_2016}}
In practice, an empirical value of $\eta$ around \SI{100}{\milli\eV} is suggested.
Other choices are legitimate like the regularizers considered by Head-Gordon and coworkers within orbital-optimized second-order M{\o}ller-Plesset theory \alert{(MP2)}, which have the specificity of being energy-dependent. \cite{Lee_2018a,Shee_2021}
In this context, the real version of the simple energy-independent regularizer \eqref{eq:simple_reg} has been shown to damage thermochemistry performance and was abandoned. \cite{Stuck_2013,Rostam_2017}

Our investigations have shown that the following energy-dependent regularizer
\begin{equation}
\label{eq:srg_reg}
	f_{\alert{\kappa}}(\Delta) = \frac{1-e^{-2\Delta^2/\alert{\kappa}^2}}{\Delta}
\end{equation}
derived from the (second-order) perturbative analysis of the similarity renormalization group (SRG) equations \cite{Wegner_1994,Glazek_1994,White_2002} by Evangelista \cite{Evangelista_2014} is particularly convenient and effective for our purposes.
Increasing $\alert{\kappa}$ gradually integrates out states with denominators $\Delta$ larger than $\alert{\kappa}$ while the states with $\Delta \ll \alert{\kappa}$ are not decoupled from the reference space, hence avoiding intruder state problems. \cite{Li_2019a}

Figure \ref{fig:H2reg_zoom} compares the non-regularized and regularized quasiparticle energies in the two regions of interest \alert{for various $\eta$ and $\kappa$ values.}
It clearly shows how the regularization of the $GW$ self-energy diabatically linked the two solutions to get rid of the discontinuities.
However, this diabatization is more or less accurate depending on \alert{(i) the actual form of the regularizer, and (ii) the value of $\eta$ or $\kappa$.}

Let us first discuss the simple energy-independent regularizer given by Eq.~\eqref{eq:simple_reg} (top panels of Fig.~\ref{fig:H2reg_zoom}).
Mathematically, in order to link smoothly two solutions, the value of $\eta$ has to be large enough so that the singularity lying in the complex plane at the avoided crossing is moved to the real axis (see Ref.~\onlinecite{Marie_2021} and references therein).
This value is directly linked to the difference in energy between the two states at the avoided crossing, and is thus, by definition, energy-dependent.
This is clearly evidenced in Fig.~\ref{fig:H2reg_zoom} where, depending on the value of $\eta$, the regularization is more or less effective.
For example, around $\RHH = \SI{1.1}{\angstrom}$ (top-left), a value of \SI{0.1}{\hartree} (green curve) is appropriate while at $\RHH = \SI{0.5}{\angstrom}$ (top-right), this value does not seem to be large enough.
Note also that $\eta = \SI{0.1}{\hartree}$ is significantly larger than the suggested value of \SI{100}{\milli\eV} and if one uses smaller $\eta$ values, the regularization is clearly inefficient.

Let us now discuss the SRG-based energy-dependent regularizer provided in Eq.~\eqref{eq:srg_reg} (bottom panels of Fig.~\ref{fig:H2reg_zoom}).
For $\alert{\kappa} = \SI{10}{\hartree}$, the value is clearly too large inducing a large difference between the two sets of quasiparticle energies (purple curves).
For $\alert{\kappa} = \SI{0.1}{\hartree}$, we have the opposite scenario where $\alert{\kappa}$ is too small and some irregularities remain (green curves).
We have found that $\alert{\kappa} = \SI{1.0}{\hartree}$ is a good compromise that does not alter significantly the quasiparticle energies while providing a smooth transition between the two solutions. 
Moreover, \alert{although the optimal $\kappa$ is obviously system-dependent}, this value performs well in all scenarios that we have encountered.
However, it can be certainly refined for specific applications. 
\alert{For example, in the case of regularized MP2 theory (where one relies on a similar energy-dependent regularizer), a value of $\kappa = 1.1$ have been found to be optimal for noncovalent interactions and transition metal thermochemistry. \cite{Shee_2021}}

To further evidence this, Fig.~\ref{fig:H2reg} reports the difference between regularized (computed at $\alert{\kappa} = \SI{1.0}{\hartree}$ with the SRG-based regularizer) and non-regularized quasiparticle energies as functions of $\RHH$ for each orbital.
The principal observation is that, in the absence of intruder states, the regularization induces an error below \SI{10}{\milli\eV} for the HOMO ($p = 1$) and LUMO ($p = 2$), which is practically viable.
Of course, in the troublesome regions ($p = 3$ and $p = 4$), the correction brought by the regularization procedure is larger (as it should) but it has the undeniable advantage to provide smooth curves.
\alert{Similar graphs for $\kappa = \SI{0.1}{\hartree}$ and $\kappa = \SI{10}{\hartree}$ [and the simple regularizer given in Eq.~\eqref{eq:simple_reg}] are reported as {\SupMat}, where one clearly sees that the larger the value of $\kappa$, the larger the difference between regularized and non-regularizer quasiparticle energies.}

As a final example, we report in Fig.~\ref{fig:F2} the ground-state potential energy surface of the \ce{F2} molecule obtained at various levels of theory with the cc-pVDZ basis.
In particular, we compute, with and without regularization, the total energy at the Bethe-Salpeter equation (BSE) level \cite{Salpeter_1951,Strinati_1988,Blase_2018,Blase_2020} within the adiabatic connection fluctuation dissipation formalism \cite{Maggio_2016,Holzer_2018b,Loos_2020e} following the same protocol as detailed in Ref.~\onlinecite{Loos_2020e}.
These results are compared to high-level coupled-cluster \alert{(CC)} calculations extracted from the same work: \alert{CC with singles and doubles (CCSD) \cite{Purvis_1982} and the non-perturbative third-order approximate CC method (CC3). \cite{Christiansen_1995b}}
As already shown in Ref.~\onlinecite{Loos_2020e}, the potential energy surface of \ce{F2} at the BSE@{\GOWO}@HF (blue curve) is very ``bumpy'' around the equilibrium bond length and it is clear that the regularization scheme (black curve computed with $\alert{\kappa} = \SI{1}{\hartree}$) allows to smooth it out without significantly altering the overall accuracy.
Moreover, while it is extremely challenging to perform self-consistent $GW$ calculations without regularization, it is now straightforward to compute the BSE@ev$GW$@HF potential energy surface (gray curve). 
\alert{For the sake of completeness, similar graphs for $\kappa = \SI{0.1}{\hartree}$ and $\kappa = \SI{10}{\hartree}$ are reported as {\SupMat}.
For $\kappa = \SI{0.1}{\hartree}$, one still has issues. 
In particular, the BSE@ev$GW$@HF calculations do not converge for numerous values of the internuclear distance. 
Interestingly, for $\kappa = \SI{10}{\hartree}$, the smooth BSE@{\GOWO}@HF and BSE@ev$GW$@HF curves are superposed, and of very similar quality as CCSD.}

%%%%%%%%%%%%%%%%%%%%%%%%%%%%%%%%%%%%%%%%%%%%%%
\section{Concluding remarks}
%%%%%%%%%%%%%%%%%%%%%%%%%%%%%%%%%%%%%%%%%%%%%%
In the present article, we have provided mathematical and physical explanations behind the appearance of multiple solutions and discontinuities in various physical quantities computed within the $GW$ approximation.
More precisely, we have evidenced that intruder states are the main cause behind these issues and that this downfall of $GW$ is a key signature of strong correlation.
A simple and efficient regularization procedure inspired by the similarity renormalization group has been proposed to remove these discontinuities without altering too much the quasiparticle energies.
Moreover, this regularization of the self-energy significantly speeds up the convergence of (partially) self-consistent $GW$ methods.
We hope that these new physical insights and technical developments will broaden the applicability of Green's function methods in the molecular electronic structure community and beyond.

%%%%%%%%%%%%%%%%%%%%%%
\section*{Supplementary Material}
\label{sec:supmat}
%%%%%%%%%%%%%%%%%%%%%%
\alert{Included in the {\SupMat} are the raw data associated with each figure as well as additional figures showing the effect of the regularizer and its parameter, with, in particular, the difference between non-regularized and regularized quasiparticle energies for \ce{H2}, and the ground-state potential energy surface of \ce{F2} around its equilibrium geometry.}

%%%%%%%%%%%%%%%%%%%%%%%%
\acknowledgements{
The authors thank Pina Romaniello, Fabien Bruneval, and Xavier Blase for insightful discussions.
This project has received funding from the European Research Council (ERC) under the European Union's Horizon 2020 research and innovation programme (Grant agreement No.~863481).}
%%%%%%%%%%%%%%%%%%%%%%%%

%%%%%%%%%%%%%%%%%%%%%%%%%%%%%%%%
\section*{Data availability statement}
%%%%%%%%%%%%%%%%%%%%%%%%%%%%%%%%
The data that supports the findings of this study are available within the article and its supplementary material.


\end{document}